\documentclass{llncs}

\pdfoutput=1

\usepackage[pdftex]{graphicx}
\usepackage{url}
\usepackage{subfigure}

\usepackage{wrapfig}

\title{Predicting Intermediate Storage Performance for Workflow Applications}
\author{Lauro Beltr\~{a}o Costa\inst{1} \and
Abmar Barros\inst{1}\inst{2} \and
Samer Al-Kiswany\inst{1} \and
Emalayan Vairavanathan\inst{1} \and
Matei Ripeanu\inst{1}
\institute{ECE Department - The University of British Columbia
\and Laborat\'{o}rio de Sistemas Distribu\'{i}dos - Universidade Federal de Campina Grande}
\email{\inst{1}\{lauroc,emalayan,samera,matei\}@ece.ubc.ca; \inst{2}abmar@lsd.ufcg.edu.br}}
\begin{document}
 \pdfoutput=1

%The command
\maketitle
%then formats the complete heading of your article. If you leave it out the work
%done so far will produce no text.

%Then the abstract should follow. Simply code
\begin{abstract}
Configuring a storage system to better serve an application is a challenging task complicated by a multidimensional, discrete configuration space and the high cost of space exploration (e.g., by running the application with different storage configurations). To enable selecting the best configuration in a reasonable time, we design an end-to-end performance prediction mechanism that estimates the turn-around time of an application using storage system under a given configuration.  This approach focuses on a generic object-based storage system design, supports exploring the impact of optimizations targeting workflow applications (e.g., various data placement schemes) in addition to other, more traditional, configuration knobs (e.g., stripe size or replication level), and models the system operation at data-chunk and control message level. 

This paper presents our experience to date with designing and using this prediction mechanism. We evaluate this mechanism using micro- as well as synthetic benchmarks mimicking real workflow applications, and a real application.. A preliminary evaluation shows that we are on a good track to meet our objectives: it can scale to model a workflow application run on an entire cluster while offering an over 200x speedup factor (normalized by resource) compared to running the actual application, and can achieve, in the limited number of scenarios we study, a prediction accuracy that enables identifying the best storage system configuration. 
\end{abstract}

\section{Introduction}

\label{sec:intro}

Assembling workflow applications by putting together standalone binaries has become a popular approach to build complex scientific applications (e.g., modFTDock \cite{Modftdock:12}, Montage\cite{Montage:03} or BLAST \cite{blast}). The processes spawned from these binaries communicate via temporary files stored on a shared storage system. In this setup, the workflow runtime engines are basically schedulers that build and manage a task-dependency graph based on the tasks' input/output files (e.g., SWIFT \cite{Swift:11}, Pegasus \cite{Pegasus:05}).

To avoid accessing the platform's backend storage system (e.g., NFS or GPFS), the shared storage system can be co-deployed on the nodes allocated to the application. Aggregating node-local resources to provide such a shared \textit{intermediate storage system} \cite{versatile,Wozniak:09} offers a number of advantages: higher performance - as applications benefit from a wider I/O channel obtained by striping data across several nodes; higher efficiency – as it improves resource utilization; incremental scalability – as it is possible to increase system capacity in small increments. This scenario also opens the opportunity for optimizing the intermediate storage system for the target workflow application: a storage system used by a single workflow, and co-deployed on the application nodes, can be configured specifically for the I/O patterns generated by the workflow (e.g., configure striping and replication to eliminate hot spots, use a data placement policy to maximize data access locality) \cite{Mosa-ccgrid:12}. 

These benefits, however, come at a price: configuring the intermediate storage system becomes increasingly complex for multiple reasons. First, the optimization techniques commonly used in distributed environments expose trade-offs that rarely exist in centralized solutions \cite{thereska:informed:06,strunk-utility}. Second, each workflow application obtains peak performance at a different configuration point, a consequence of different I/O patterns \cite{ursaminor}. Third, depending on the context, there are multiple metrics of interest to optimize: e.g., response time, throughput, energy, and, increasingly common in cloud computing environments, the cost of resources. 

To illustrate these points, consider the following optimization techniques and their trade-offs. Data striping may accelerate data access, yet it decreases reliability and requires more resources. Increasing the number of storage nodes can avoid access bottlenecks, yet it may expose scalability bottlenecks. Higher redundancy levels (through replication or erasure codes) may accelerate data access and increase reliability, yet they also require more resources and complex consistency protocols. Finally, different data placement and/or caching policies are beneficial to different workloads.

\begin{figure}[htb]
\centering
\includegraphics[width=0.5\linewidth]{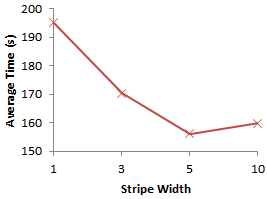} 
\caption{\textit{Different storage system configurations deliver different performance and the choice of the optimal configuration point is not intuitive.} To demonstrate that this holds for workflow applications, we have executed the Montage workflow \cite{Montage:03} on top of MosaStore \cite{versatile} storage system with different storage configurations deployed over the Grid'5000 platform. We executed the same workload for several values of one configuration knob: \textit{stripe width}. As we increase stripe width, we also increase the number of storage nodes to support it. For a low stripe width, performance is low due to congestion at the storage nodes and, for higher values, performance decreases due to connection handling and metadata access overheads. A user seeking the optimal point would pick a stripe width value close to five, a non-obvious choice. Other case studies can be found in \cite{Mosa-ccgrid:12}. }

\label{fig:montage}
\end{figure}

Past work and our own experience support these assertions. They show that different applications achieve their best performance when using different storage configurations \cite{versatile,ursaminor}. Also, different applications even benefit from different combinations of storage optimization techniques \cite{Mosa-ccgrid:12,erss2011}, and that the choice of the optimal configuration point is not intuitive (Figure \ref{fig:montage}). 

In this scenario, the role of the application administrator/user becomes non-trivial: in addition to being in charge with running the workflow application, she also becomes responsible for choosing the appropriate intermediate storage system configuration to maximize performance. Providing support for this activity is the focus of our project \cite{grid2010}.

% Stripe	1	3	5	10
% Bene				
% 1	195.0885	#NAME?	#NAME?	#NAME?
% 3	158.804	170.333	#NAME?	#NAME?
% 5	153.842	152.944	156.1595714	#NAME?
% 10	155.259	151.657	153.6348	159.8885

\textbf{The Problem.} Configuring the intermediate-storage system to achieve high performance (e.g., in terms of application turnaround time, storage footprint, energy consumption, or financial cost), involves choosing a set of storage system configuration parameters (such as stripe width, data placement policy, and replication level).

\textit{Manually} fine-tuning the storage system configuration parameters is undesirable for multiple reasons. First, the user lacks a deep understanding of how different configuration choices affect the system's performance. Second, the workload may vary: new application versions may make one-time tuning obsolete. Finally, performance tuning is time-consuming due to the large configuration space to consider.

\textit{This paper presents our progress to date on designing and evaluating a performance prediction mechanism for object-based storage systems}. Given a storage system configuration, an application I/O profile, and a characterization of the deployment platform based on a simple system identification process (e.g., storage nodes service time, network characteristics), the mechanism predicts the total application turnaround. 

This approach can support four important tasks: First, \textit{autotuning}, a software tool that relies on the proposed mechanism can enable efficiently configuring the storage system \cite{grid2010,thereska:informed:06,strunk-utility}, through exploring the configuration space without actually running the application. Second, \textit{resource provisioning}, it can inform the decision of how many resources to allocate to the application to achieve a certain performance level. Third, \textit{new technology evaluation}, it can estimate the application performance on hardware that has not yet been procured (e.g., to help answer the question: \textit{what} would be the performance improvement \textit{if} we used SSDs?). Finally, the mechanism can help explore the impact of configuration choices in situations where direct measurement is difficult or requires specialized infrastructure (e.g., energy consumption). 

This paper focuses on predicting the application turnaround time, but we note that the model and approach presented apply readily to other optimization metrics.

\textbf{The contributions of this paper} lay over multiple axes: 

\begin{itemize}

\item Synthesizes the key requirements for a prediction mechanism that will be useful in practice (\S\ref{sec:requirements}).

\item Proposes a simple queue-based model for distributed, object-based storage systems (\S\ref{sec:model}). This model, at the base of the prediction mechanism, is: generic (i.e., covers the traditional design for object-based datastores), and homogeneous (all nodes are based on the same model). 

\item Proposes a system identification procedure to seed the model that is simple, lightweight, effective, and does not require system changes to collect monitoring information (\S\ref{sec:seeding}).

\item Presents preliminary experience with using this prediction mechanism (\S\ref{sec:evaluation}) in differentiating between system configurations and identifying the best configuration for a certain workload. Also, the paper discusses our experience (\S\ref{sec:discussion}) to date with using the prediction mechanism beyond our original design goal: we have used the performance prediction to better understand and debug a distributed storage system that our group develops.
\end{itemize}
\section{The Design of a Performance Estimation Mechanism}

\label{sec:solution}

Building an accurate and scalable distributed system performance predictor is challenging, which resulted in some simulation-based solutions instead of pure analytical models. At the one end of the design spectrum current practice (e.g., NS2 simulator \cite{ns2}) suggests that while simulating a system at low granularity (e.g., packet level simulation in NS2) can provide high accuracy, the complexity of the simulation model and the number of events generated make simulating large-scale systems infeasible. At the other end, coarse grained simulations (e.g., \cite{SimGrid,PeerSim}) can scale often at the cost of lower accuracy %[Add some example here.]. 

Similarly, the proposed mechanism relies on a simulation-based approach. We target a scalable and adequately accurate performance prediction. While we simulate the storage system at chunk granularity to increase the accuracy, we exploit the following two observations to reduce simulation complexity and increase its scalability: First, as the goal of the simulation is to support configuration choice for a specific workload, achieving perfect accuracy is less critical (\S\ref{sec:evaluation}). Second, the focus is on the workload characteristics of workflow applications: relatively large files, single-write many reads and specific data access patterns. These observations enable us to reduce the simulation complexity by not simulating in detail some of the control paths that do not significantly impact accuracy (e.g., the chunk transfer time is dominated by the time to send the data, consequently not accounting the time of the acknowledgment messages or the metadata message transfer will not tangibly impact accuracy).

 These design decisions allowed us to simulate storage operations of runs of an entire application on a cluster in an order of magnitude faster than the actual run (\S\ref{sec:evaluation}).

The proposed mechanism uses a queue-based storage system model for the system components’ operations and their interactions. The model requires three inputs from the user: the storage system configuration, a workload description, and the performance characteristics of storage system components (i.e., system identification). The simulator instantiates the storage system model with the specific component characteristics and configuration, and simulates the application run as described by the workload description.

This section discusses the requirements for a practical performance prediction mechanism (\S\ref{sec:requirements}) and presents the key aspects of the object-based storage system architecture modeled (\S\ref{sec:mosa}). Then, it focuses on the proposed solution: it presents the model (\S\ref{sec:model}), its implementation (\S\ref{sec:implementation}), the system identification process to seed the model (\S\ref{sec:seeding}), and an overview of the workload description (\S\ref{sec:workload}). 
\subsection{Solution Requirements}

\label{sec:requirements}

A practical performance prediction mechanism should meet the following, partially conflicting, requirements:

\begin{itemize}

\item \textbf{Accuracy.} The mechanism should provide adequate accuracy. Of course better accuracy is desirable; however, in the face of practical limitations to achieve perfect accuracy, we note that there are decreasing incremental gains for improved accuracy in practical settings. For example, to support decisions about configuration choices a predictor only needs to correctly estimate their relative performance or trends of changing a configuration parameter. Even more, if two configurations offer near performance, their relative predicted performance is less important as long as the prediction mechanism places their performance as similar.

\item \textbf{Scalability and Response Time.} The predictor should enable the quick exploration of the configuration space. To this end, the mechanism should offer performance predictions quickly and scale across at least two dimensions: (i) it should scale with the system size and be able to model large systems; and (ii) it should scale with the I/O intensity and be able to model I/O intensive applications. 

\item \textbf{Usability and Generality.} The predictor should not impose a burdensome effort to be used. Specifically, the bootstrapping/seeding process should be simple and it should not require storage system redesign (or a particular initial design) to collect performance measurements. Additionally, ideally the prediction mechanism should model a generic object-based distributed storage design and using it should not require in-depth knowledge of storage system protocols and architecture. 

\item \textbf{Ability to explore \textit{``what-if''} scenarios.} A prediction mechanism should be able to support exploring hypothetical scenarios, such as scenarios that assume new/different hardware configurations (e.g., usage of SSDs). We note that there are two main categories for the models at the foundation of all performance prediction mechanisms: explanatory and agnostic models. The explanatory models try to mimic the key components of the system and their interaction at various levels of accuracy and granularity. The agnostic model aims to predict the output metric of interest being completely oblivious of system internals (e.g., a neural network based models would fit in this category). Supporting \textit{``what-if''} scenarios exploration requires an approach based on an explanatory model.

\end{itemize}
\subsection{Object-based Storage System Design}

\label{sec:mosa}

We focus on a widely-adopted object-based storage system architecture (such as that adopted by GoogleFS ‎\cite{ghemawat2003google}, PVFS \cite{pvfs}, MosaStore \cite{versatile}, and UrsaMinor \cite{ursaminor}). This architecture includes three main components: a centralized metadata manager, storage nodes, and a client-side system access interface (SAI). The manager maintains the stored files' metadata and system state. To speed up data storage and retrieval, the architecture employs striping ‎\cite{zebra95}: files are split into chunks stored across several storage nodes. Client SAIs implement data access protocols after they interact with the manager that stores data placement information. 

\textit{Data placement}. The default data placement generally adopted is round-robin: when a new file is created on a stripe of \textit{n} nodes the file's chunks are placed in a round-robin fashion across these nodes. Additionally, data placement policies that optimize for a specific application access patterns have seen higher adoption \cite{Mosa-ccgrid:12,pnfs,Zhang:2012:DAD:2388996.2389112}. For instance, the following data placement policies are used to optimize for the workflow applications' data access patterns: local, co-locate and broadcast (detailed in \S\ref{sec:evaluation}).

\textit{Replication}. Replication is often used to increase reliability or to improve access performance. Data is replicated when new data is stored in the system, consequently, while a higher replication level will reduce contention on the node storing a popular file, it will increase the file write time and the storage space consumption.

We explore the accuracy of the prediction mechanism assuming that the stripe width, replication level, and data placement policy are configurable as suggested in \cite{Mosa-ccgrid:12,ursaminor,versatile}. Our approach can be extended to support other configuration parameters.
\subsection{System Model}

\label{sec:model}

The process of developing the prediction mechanism involves several trade-offs between simplicity and accuracy. The solution space is bounded by the requirements listed in \ref{sec:requirements}: we avoid changes in the storage system itself, and make the model and the system identification as simple as possible. 

\begin{figure}[htb]
\centering
\includegraphics[width=0.75\linewidth]{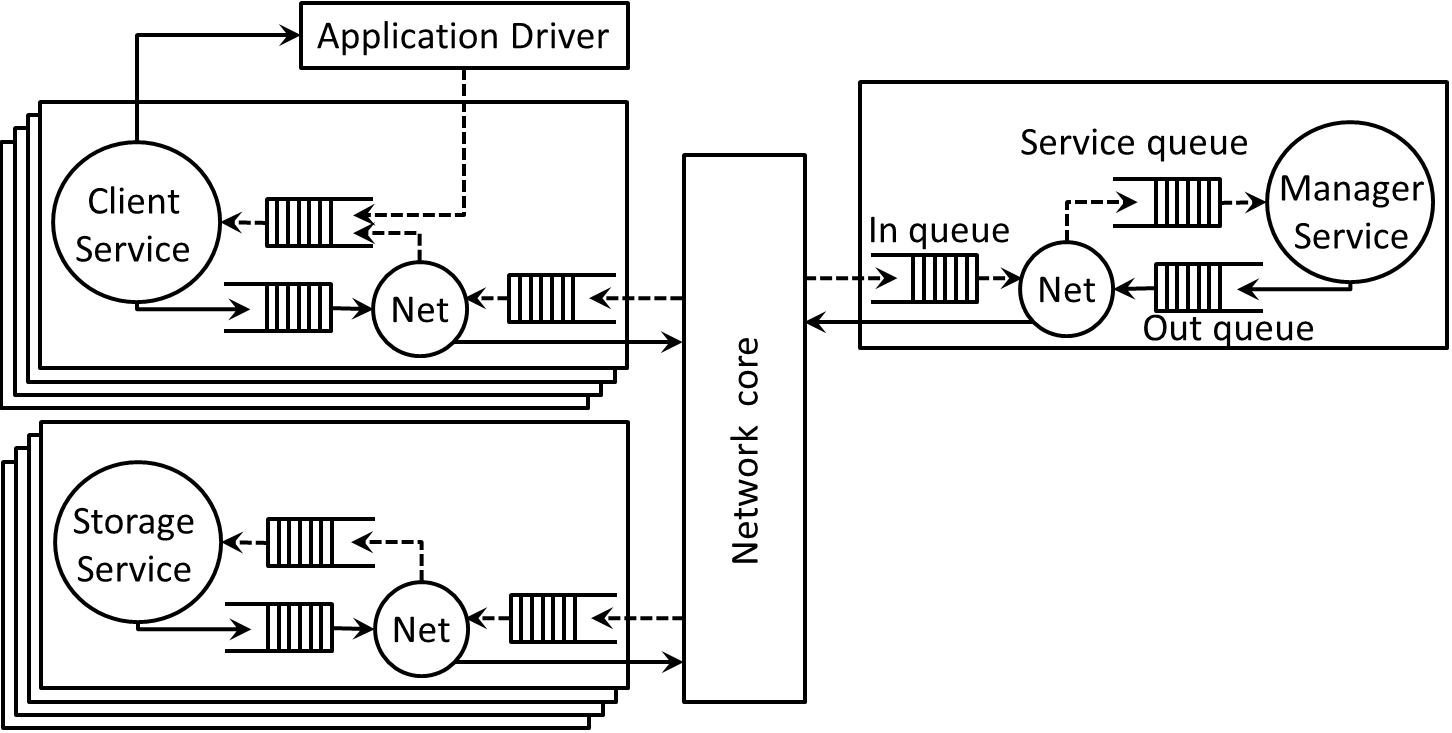} 
\caption{Queue-based model of a distributed storage system. Each component (manager, client component, and storage component) has a single system service that processes requests from its queue. Additionally, each host has a network service with an in- and out- queue. The network core connects and routes the messages between the different components in the system and can model network latency and contention at the aggregate network fabric level. Solid lines show the flow going out from a storage system component while dashed lines show the in-flow path.}
\label{fig:model}
\end{figure}

We use a queue-based model. All machines are modeled similarly, regardless of their specific role in the system (Figure \ref{fig:model}): each machine hosts a network component and can host one or more system components (each modeled as a service with its own queue).

A system service and its queue represent a specific functionality in the system. That is, the \textit{manager} component is responsible for storing files' and storage nodes' metadata. The \textit{storage} component is responsible for storing and replicating data chunks. Finally, the \textit{client} component receives the read and write operations from the application, implements, at the high-level the storage system protocol by sending control or data requests to other services, and once a storage operation is terminated it communicates again with the application. Each of these components is modeled as service that takes requests from its queue (fed by the network service or by the application for the client service) and sends responses back through the network service (or directly to the application, again for the client). 

The network service and its in- and out- queues model the network-related activity of a host. Key here is to model network-related contention while avoiding modeling the details of the transport protocol (e.g., dealing with packet loss, connection establishment and teardown details). The requests in the out-queue of a network component are broken in smaller pieces that represent network frames and sent to the in-queue of the destination host. Once the network service processes all the frames of a given request in the in-queue, it assembles the request and places it in the queue of the destination service.

The system services can be collocated on the same host (e.g., the client and storage services running on the same host). In this situation, requests between collocated services also go through the network, but have a faster service time than remote requests - representing a loopback data transfer (\S\ref{sec:seeding}).

Space limitations prevent us from presenting the full details of the model. As a rule, we uniformly model the data path at chunk-level granularity, and the control paths at a coarser granularity: modeling only one control message to initiate a specific storage function while an implementation may have multiple rounds of control message exchanges. We used our own experience with designing and implementing an object-based storage system \cite{mosastore-website}.

\subsection{Model Implementation}

\label{sec:implementation}

We have implemented the above model as a discrete-event simulator in Java. The simulator receives as inputs: a summarized description of the application workload (described in \S\ref{sec:workload}) and a description of the deployed system which has two parts. The first part describes the system-wide configuration parameters (currently, replication level, stripe-width, chunk size, and data-placement system-wide) and the details of the system: number of hosts, number of storage nodes and clients, whether storage and clients nodes are collocated on the same hosts. The second part characterizes the performance of system services: service times for network, client, storage, and manager services (the process of identifying these values is described in \S\ref{sec:seeding}). 

Once the simulator instantiates the storage system, it starts the application driver that processes the application workload. The driver reads the description of the application workload, creates the corresponding events (e.g., read from file $x$ at offset $y$, $z$ bytes) and places them in the client service queue. File-specific configuration (as proposed by \cite{ursaminor,Mosa-ccgrid:12}) is described as part of the application workload description since it depends on the file that application reads/writes.

As in a real system, the manager component maintains the metadata of the system (i.e., implements data placement policies by returning free chunks when requested by write operations, and keeps track of file to chunk mapping and chunk placement). To make the process clearer, consider the following example for a write operation where a client module processes a file write event. First, the client contacts the manager asking for free space. The manager replies specifying a set of free chunks and their storage services to be used during this write. Then, the client requests each storage service to store chunks in a round-robin fashion. After processing a request to store a chunk, a storage service replies to the client acknowledging the operation success. After sending all the chunks, the client sends to the manager the chunk-map (where each chunk is stored). Then, once the manager acknowledges, the client returns success to the application driver. In total the write operation generates two requests to the manager and one request per chunk to the storage nodes. 

Note that the specific set of storage services returned by the manager in the beginning of a write operation depends on the data placement policy used. A typical set is composed of stripe-width of storage services, but it can be composed of just one, e.g., when a local data placement is desired. 

To model per-file optimizations, the client can overwrite system-wide configurations by requesting the manager to provide support for a specific data placement scheme. For example, the client may require that a file is stored locally, that is, on a storage service that is located on the same host. In this case, the manager attempts to allocate space on that specific storage service for that write operation (as opposed to striping the data across multiple storage services). The file-specific data placement policy is part of the workload description. 

All communication among the system services uses the network. Each network request has its destination address, which is used by the simulator to determine which network queue should receive a packet. 

Currently, the simulator reports the time spent, data transferred and storage used per each read or write. Additionally, one may request to collect aggregated information for specific points of the simulation. %(e.g., all files of a given stage of the application were written).
\subsection{System Identification}

\label{sec:seeding}

To instantiate the storage system model, one needs specify the number of storage and client components in the system, and define the service times for the network ($\mu^{net}$) and the system components (storage - $\mu^{sm}$, manager - $\mu^{ma}$, and client - $\mu^{cli}$). 

The system identification process is automated with a script as follows. First, to measure the service time per chunk/request $T^{net}$), a script runs a network throughput measurement utility tool (e.g., \textsf{iperf} \cite{iperf}, ), to measure the throughput of both: remote and local (loopback) data transfers. Second, this script measures the time to read/write a number of files to identify client and storage service time per data chunk. To this end, the system identification script deploys one client, one storage node and the manager on different machines, and writes/reads a number of files. For each file read/write the benchmark records the total operation time. At the end of its execution, the script computes the average read/write time $T^{tot}$. The number of files read/wrote is set to achieve 95\% confidence intervals with $\pm5\%$ accuracy according to the procedure described in \cite{jain91}.

The operation total time ($T^{tot}$) includes the client side processing time ($T^{cli}$), the storage node processing time ($T^{sm}$), the total time related to the manager operations ($T^{man}$) , and the network transfer time ($T^{net}$). The network service time for the network ($\mu^{net}$) is based on a simple analytical model based on network throughput and proportional to the amount of data to be transferred in a packet.

To isolate just $T^{cli} + T^{man}$, the script runs a set of reads and writes of 0-size. This forces a request to go through the manager, but it does not touch the storage module. Since decomposing $T^{cli}$ and $T^{man}$ is not possible without probes in the storage system code, we opted to associate the $T^{cli} = 0$ and associate the whole cost of 0-size operations to the manager. While iperf can estimate $T^{net}$, and the script can infer $T^{cli} + T^{man}$, and therefore $T^{sm} = T^{tot} - T^{net} - T^{man}$. To obtain the service time per chunk, the times are normalized by chunk size. Therefore, $\mu^{sm} = \frac{T^{sm}}{chunkSize}$.
\subsection{Workload Description}

\label{sec:workload}

The simulator takes as an input a description of the workload to be simulated. The workload description contains two pieces of information: per client I/O operations trace (i.e., open, read, write, close calls with the call details: timestamp, operation type, size, offset, and client id), and a files' dependency graph (capturing the operation dependency). The client traces can be obtained by running and profiling the application. The storage system logs often already provide these traces. Generating the file dependency graph can either be provided by the workflow scheduler (e.g., Swift \cite{Swift:11}), by an expert user or automatically extracted from log files. Automating the extraction of the file dependency graph and client traces from storage system logs is an ongoing effort at our research group and is out of the scope of this paper.
\section{Evaluation}

\label{sec:evaluation}

This section aims to evaluate the mechanism's prediction accuracy and, more importantly, to demonstrate through a set of experiments the mechanism's ability to support correctly identifying the best configuration for a specific application pattern. To this end, we use a set of micro- and synthetic benchmarks, and a real application.. The microbenchmark-based evaluation compares the runtime of a simple workload of a single operation (read or write of a single file) to the predicted runtime provided by the simulator under a spectrum of configurations. The evaluation demonstrates that the mechanism can accurately predict in these simple scenarios the performance of the operations in all possible system configurations (varying stripe width and replication level). The synthetic benchmarks are designed to mimic real workflow application access patterns \cite{Mosa-ccgrid:12}. The goal is to evaluate the mechanism's ability to predict performance under more complex system interactions that resemble the application ecosystem, we target and the storage system optimizations used in this context. Finally, we use BLAST \cite{blast} as an example of a real application.

\textbf{Storage system.} The storage system used is MosaStore. The storage nodes are backed up by RAMDisk. We choose to experiment with RAMDisks as they are frequently used to support workflow applications: it offers higher performance and are the only option in some supercomputers that do not have spinning disks (e.g., IBM BG/P machines). We briefly discuss a disk based evaluation in \S \ref{sec:discussion}.

\textbf{Deployment platform.} We use a testbed of 20 machines each with an Intel Xeon E5345 4-core, 2.33-GHz CPU, 4GB RAM, and 1-Gbps NIC. One machine runs the MosaStore manager while the other 19 machines each run both a storage node and a client access module.

\textbf{System identification.} The simulator is seeded according to the procedure described in \S\ref{sec:seeding}.
\subsection{Microbenchmarks}

\label{subsec:microbenchmarks}
We first focus on simple experiments composed of a set of read or write operations of 100MB files, triggered from one client. We used this experiment to guide the modeling/implementation effort. To explore different configuration choices, we used: \textit{stripe width} and \textit{replication level} (with values from 1 to 5 for each), producing a total of 25 configuration scenarios. Figure \ref{fig:relativerror} shows the relative error of the predicted runtime for the write microbenchmark (similar results were obtained for reads). For most of the cases, predicted performance is within 10\% of the actual. 

\begin{figure}[htb]
\centering
\includegraphics[width=0.4\linewidth]{relative-error-writes} 
\caption{Relative error for predicted performance. Values greater than 0 indicate that actual times are longer than predicted. We present averages over 30 runs. For most scenarios error within 5\%, for all scenarios it is within 20\% (Update Figure)}
\label{fig:relativerror}
\end{figure}

\subsection{Synthetic Benchmarks for Workflow Patterns}
\label{sec:eval:synt}

\begin{figure}[htb]
\centering
\includegraphics[width=0.6\linewidth]{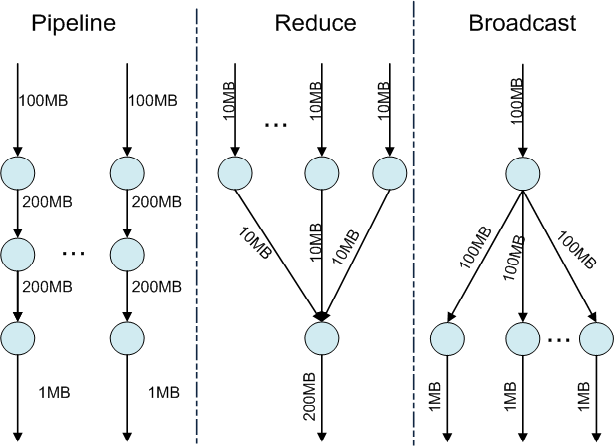} 
\caption{Pipeline, Reduce, and Broadcast benchmarks. Nodes represent workflow stages and arrows represent data transfers through files. The file sizes represent the \textit{medium} workload. Files in the \textit{large} workload are 10x larger.}
\label{fig:patterns}
\end{figure}

The microbenchmarks provide important data to understand the basic system operation and the impact of changes in the main knobs. They do not, however, capture the behavior of the system with multiple clients, the interaction among multiple applications, or the impact of data-placement policies designed to support workflow applications. To evaluate the accuracy of the prediction mechanism in more scenarios, we use synthetic benchmarks that mimic common data access patterns that exist in workflow applications \cite{Mosa-ccgrid:12}. Specifically, this section focuses on pipeline, reduce, and broadcast patterns (Figure \ref{fig:patterns}). These are among the patterns uncovered by studying over 20 scientific workflow applications by Wozniak et al. \cite{Wozniak:09}, Shibata et al. \cite{Shibata:10}, and Bharathi, et al. \cite{Bharathi:08}). 

\textbf{Experimental setup}. We use the MosaStore setup described above. We use the \textit{DSS} label (from generic Distributed Storage System) for experiments where we use MosaStore default configuration: client and storage modules run on all machines, client stripes data over all 19 machines, and there is no data-access pattern optimization enabled. We use the \textit{WASS} label (Workflow Aware Storage System) when the system configuration is optimized for a specific access pattern (including data placement, stripe width or replication) \cite{Mosa-ccgrid:12}. All WASS experiments assume data location aware scheduling: for a given compute task, if all input file chunks exist on a single storage node, the task is scheduled on that node to increase access locality.

The goal of showing results for two different configurations choices is two-fold: (i) demonstrate the accuracy of the predictions for two different scenarios, and (ii), most important, show that the predictions correctly indicates which configuration is the best. To understand the impact of the data size, for each benchmark, we use (where possible) two workloads labeled as the \textit{medium} and, a 10x larger, \textit{large} workload. We omit results for a \textit{small} workload, which is 10x smaller than medium, because it exhibits a high variability and similar performance between different configurations \cite{Mosa-ccgrid:12}. 

For actual performance, the figures show the average turnaround time and standard deviation for 15 trials. It is enough to guarantee a 95% confidence level.

\textbf{Pipeline benchmark}. A set of compute tasks are chained in a sequence such that the output of one task is the input of the next task in the chain (Figure \ref{fig:patterns}). A pipeline-optimized storage system will store the intermediate pipeline files on the storage node co-located with the application. Later, the workflow scheduler places the task that consumes the file on the same node, increasing data access locality. Here, 19 application pipelines run in parallel and go through three processing stages that read input from the intermediate storage and write the output to the intermediate storage. (We present only results for the medium workload as the large workload does not fit in the RAMdisk of the machines of our testbed). 

\begin{figure}[htb]
\centering
\includegraphics[width=0.4\linewidth]{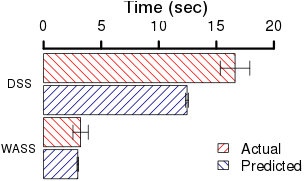} 
\caption{Actual and predicted performance for the pipeline benchmark and medium workload. Error bars show the standard deviation.}
\label{fig:pipeline-medium-accuracy}
\end{figure}

\emph{Evaluation results.} Figure \ref{fig:pipeline-medium-accuracy} shows the evaluation results. The simulator produces estimates equivalent to actual results for the optimized configuration (WASS). For no optimization (DSS), the prediction is 16\% smaller (standard deviation considered). Note that for a case with default data placement policy, all clients stripe (write) data to all machines in the system; similarly, all machines read from all others. This creates, a complex interaction among all components in the system and some retries due to connection timeouts caused by network congestion which, we believe, is the source of the prediction inaccuracy. 

\textbf{Reduce or Gather benchmark}. A single compute task uses input files produced by multiple tasks. Real-world situations that generate this pattern include a task that checks the results of other tasks executed in parallel for a convergence criterion, or a task that calculates summary statistics from the output of many tasks. A possible data placement optimization is the use of collocation i.e., placing all these input files on one node and expose their location, which will later be used by the scheduler to run the reduce task on that machine. In the benchmark, 19 processes run in parallel on different nodes, consume an input file, and produce an intermediate file. In the next stage of the workflow, a single process reads all intermediate files and produces the reduce-file. Data sizes are indicated in Figure \ref{fig:patterns}. In this scenario, for WASS configuration, the collocation optimization is enabled for the files used in the reduce stage, for the remaining files the locality optimization is enabled.

\begin{figure}
\begin{center}
\mbox{
\subfigure[Medium]
{\includegraphics[width=0.4\linewidth]{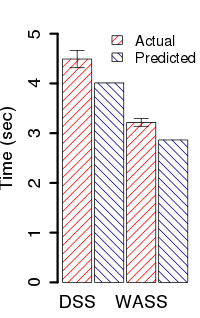}
\label{fig:reduce-medium-accuracy}}
\subfigure[Large]
{\includegraphics[width=0.4\linewidth]{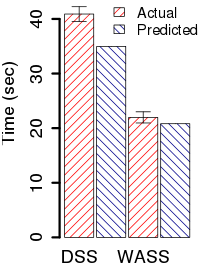}
\label{fig:reduce-large-accuracy}}
\subfigure[Large per Stage for WASS]
{\includegraphics[width=0.2\linewidth]{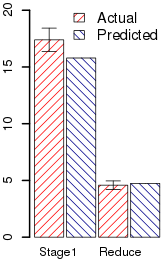}
\label{fig:reduce-large-stage}}

}
\caption{Actual and predicted performance for the reduce benchmark for the medium, large workloads, and per stage for large workload. Error bars show the standard deviation.}
\label{fig:reduce-ram}
\end{center}
\end{figure}

\emph{Evaluation results.} Similar to the pipeline benchmark, predictions for the reduce benchmark are within 20\% of the actual performance (Figure \ref{fig:reduce-ram}) and, more importantly, they capture the relative improvements that pattern-specific data placement policies policy can bring. 
We note that Figure \ref{fig:reduce-large-accuracy} captures the behavior of a heterogeneous scenario: We used a faster machine with a larger RAMDisk to run the reduce stage since the RAMDisk of the typical machine in the testbed is too small. Despite heterogeneity, the predictor captures the system performance with accuracy similar to a homogeneous system.

When the collocation and locality optimizations are not enabled, the challenge of capturing exactly the system behavior is similar to the pipeline case: capture the complex interactions among all machines in the system. When the specific data placement is enabled though, the challenge is different: there is a high contention created by having several clients writing to the same storage machine (the one that performs the reduce phase). Figure \ref{fig:reduce-large-stage} shows the results per-stage for the two stages of the large workload.

\textbf{Broadcast benchmark}. A single task produces a file that is consumed by multiple tasks. In this benchmark, 19 processes running in parallel on different machines consume a file that is created in earlier stage by one task. A possible optimization for this pattern is to create replicas of the file that will be consumed by several different tasks. Data sizes are indicated in Figure \ref{fig:patterns}.

\begin{figure}[htb]
\centering
\includegraphics[width=0.4\linewidth]{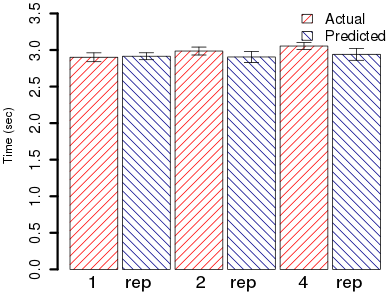} 
\caption{Actual and predicted performance for broadcast benchmark and medium workload. The experiment uses the WASS system while varying the replication level. Error bars show the standard deviation.}
\label{fig:broadcast-medium-accuracy}
\end{figure}

\emph{Evaluation results.} Figure \ref{fig:broadcast-medium-accuracy} shows the results for broadcast pattern with medium workload with the WASS system configured with 1, 2, or 4 replicas (the large workload shows a similar trend and we omit it here). For this benchmark all predictions matched the actual results: predictions were inside the interval of mean of actual $\pm$ standard deviation, just 1-2\% difference from the mean. This experiment highlights an interesting case for the predictor. According to the structure of the pattern and the results reported in \cite{Mosa-ccgrid:12} (admittedly on a slightly different setup), creating replicas would improve the performance of the broadcast pattern. The results, however, show that creating replicas does not really help here. This happens because striping to many machines already avoids the contention of a single node holding the file. So, although creating replicas can alleviate the number of accesses to a given machine (since chunks are read in sequence), this gain is not paid off by the overhead of creating a replica. More importantly in our context, this is another situation where the predictor captures the impact of different configurations, showing, in this case that they are equivalent and the user can stick with one replica and save storage space.

\textit{\textbf{Synthetic Benchmark Evaluation summary}.} The evaluation with the microbenchmark and synthetic benchmarks demonstrates that a simulator can rely on proposed prediction approach to simulate complex system interactions generated by an application run on a complete cluster. Our approach leads to errors of 6\% in average, lower than 9\% in 90\% of the studied scenarios and within 20\% in the worst case. More importantly, the mechanism correctly differentiates between the different configurations and could support choosing the best configuration for each scenario in our evaluation. Finally, simulating the system takes roughly 10x less time and uses just one machine (200x fewer resources) than the actual execution of the benchmarks.

\subsection{Real Application}
\label{sec:eval:blast}
This section present the evaluation for prediction accuracy with an entire real application (BLAST). BLAST \cite{blast} is a DNA search tool for finding similarities between DNA sequences. Each node in the cluster receives a set of DNA sequences as input (2 DNA queries in our evaluation with total size of 5.6KB) and each node searches, independently, the same database (size of 1.7GB in our evaluation) for matches with the input sequence, and writes the result (size of 82KB) to the storage system. A possible optimization for BLAST is to replicate the database file (broadcast pattern). We used replication levels of 1, 2, 4, 8, and 16. Finally, we note that, although BLAST is less I/O intensive than the synthetic benchmarks, it generates an intensive I/O workload: 550k I/O operations for this experiment.

\begin{figure}[htbp]
\centering
\includegraphics[width=0.5\linewidth]{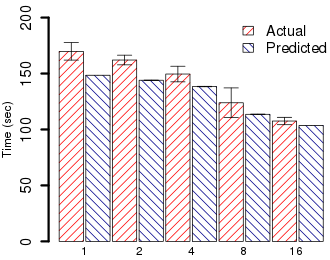}
\caption{Average of the actual and predicted performance for the BLAST application.
}
\label{fig:blast}
%\end{wrapfigure}
\end{figure}

\emph{Evaluation results.} Figure \ref{fig:blast} compares the actual and predicted performance for the BLAST workflow while varying the replication level. The results show a maximum prediction error of 10\% of the actual performance. This highlights the ability of our approach to accurately predict performance of a complex application workload compromising thousands of IO operations and to help choosing a configuration.

%\begin{wrapfigure}{r}{0.35\textwidth}
\section{Related Work}

\label{sec:related}

This section describes briefly past work on storage system performance prediction and how it differs from our work.
% Add more references

Past work used model-driven analysis to estimate performance of storage systems. For instance, Ergastulum \cite{ergastulum} targets centralized storage solution based on one enclosure to recommend an initial configuration of the system, and Hippodrome \cite{Hippodrome} relies on Ergastulum to improve the configuration based on online monitoring of the workload. The proposed predictor also support similar task. However, we consider a distributed system rather centralized storage, which brings more complex interaction among the components of the system and more configuration options.

Similar to this work, Thereska et al. \cite{thereska:informed:06} proposed a predictor mechanism for a distributed storage system. The main difference to our work is the fact that their model is much more detailed (e.g., it model CPU and network adapter). To provide such information, they propose Stardust \cite{stardust} a detailed monitoring information system that required changes to the storage system and kernel modules to add monitoring points. This approach enabled their predictor to achieve prediction within 20\% of the actual predictions depending on the workload. Our approach have achieved similar accuracy on our target workload. 

An important difference to past work on storage systems performance is our focus on a whole workflow application and the potential interaction among the workflow's phases instead of the average performance for a batch of operations or predicting performance of the system from the perspective of just one client.
 
%- Compare to Zhao’s work, they do target workflow.
Recently, Z. Zhang et al. has proposed an approach to determine the storage bottleneck for a given many-tasks application (a class of workflow) based on a set of benchmarks and executions of the the application. The approach we propose enable a richer exploration of the system by a lower cost since the predictor is able to estimate performance of a scenario that add or reduce resources and change the configuration without requiring new runs of the application of the benchmarks.
\section{Discussion and Summary}

\label{sec:discussion}

This paper makes the case for a prediction mechanism to support the task of configuring an intermediate storage system. We focus on predicting the performance of workflow applications when running on top of an intermediate object-based storage system. Specifically, we propose a solution based on a queue-based model with a number of attractive properties: a generic and uniform system model; supported by a simple system identification process that does not require specialized probes or system changes to perform the initial benchmarking; with a low runtime to obtain predictions; and, finally, with adequate accuracy for the cases we study. 

We highlight that this is an ongoing work. The discussion below aims to clarify our understanding of the limitations of our work and the lessons we have learned during this exercise so far, and to sketch our future work. 

\vspace{+6pt}

\textbf{What are the main sources of inaccuracies?}
Currently, there are sources of inaccuracies at multiple levels: First, the model does not capture all the details of the storage system (e.g., support services like garbage collection or storage node heartbeats; the control paths are simplified to match what we believe generic object-based storage would do - while we know that a FUSE-based implementation would need more complex control paths; we model all control messages as having the same size) and the environment (e.g., contention at the network fabric level or scheduling). Second, we constrain and simplify system identification even further at the cost of additional accuracy loss. Third, we do not model the infrastructure in detail (e.g., we do not model the network protocols or the spinning disks). Finally, so far the application driver uses an idealized image of the workflow application (e.g., all pipelines are launched in the simulation exactly at the same time while in the experiments on real hardware coordination overheads make them slightly staggered). We believe the latter one is the main reason of current inaccuracies in the system and should be address by a richer workload description.

\vspace{+6pt}

\textbf{What is the accuracy when the intermediate storage is deployed on spinning disks?}
So far, we have focused on predicting performance when intermediate storage is deployed over RAMDisks for two reasons: This is a common setup on large systems (as some, do not even have spinning disks) and the individual performance of RAMdisks is simpler to predict (the service time for spinning disks is history dependent due to cache behavior and position of disk head). The storage service we use does not model history-dependent behavior, thus we expect lower accuracy predictions when the system is deployed over spinning disks. (This can be fixed, by using a more sophisticated model of the storage service and it is part of our future work). 

% We have evaluated, however, how the current (unchanged) model performs when using spinning disks for the simple synthetic benchmarks used in the evaluation section. Figure \ref{fig:reduce-hdd} shows the results for the reduce pattern when using the medium and large workloads. The key observation here is that, although prediction accuracy is lower, predictions are good enough to make the correct choice between DSS and WASS, that is, the choice on whether to use the data co-placement optimization for each of the workloads (note that this optimization is beneficial in one case and it detrimental in the other one). Pipeline benchmark on HDDs shows similar results to RAMdisks (Figure \ref{fig:pipeline-medium-accuracy}).

\begin{figure}
% \centering
\begin{center}
\mbox{
{\includegraphics[width=0.25\linewidth]{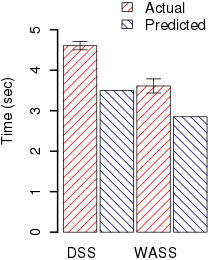}
\label{fig:reduce-medium-accuracy-hdd}}
{\includegraphics[width=0.25\linewidth]{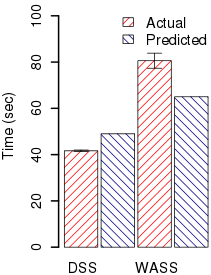}
\label{fig:reduce-large-accuracy-hdd}}
}
\caption{Actual and predicted performance for the reduce benchmark on HDD - medium(left) and large (right) workload. The results for WASS is this experiment are not directly comparable to the corresponding ones for spinning disks as we use a slower node for data co-placement, keeping the system homogeneous.}
\label{fig:reduce-hdd}
\end{center}
\end{figure}

\vspace{+6pt}

\textbf{Can the performance prediction mechanism support the \textit{development process} of a storage system? Do you have specific experience with using the mechanism in this context? }
One of the lessons we have learned so far is the utility of the mechanism to support development of the storage system itself. Back of the envelope calculations are a common mechanism to evaluate expected performance bounds for a given system. The predictor takes this a step further and is useful in complex scenarios where back of the envelope estimates were intractable. Not only developers can use it to evaluate the potential gains of implementing a new complex optimizations or to study the impact of faster network and nodes, but also the mechanism can be used as a baseline to detect performance anomalies. 

More concretely, we have encountered a number of situations where the predicted and actual performance differed significantly. In some cases these highlighted simplifications in the model or in our simulator. But, more importantly, there were cases that highlighted complex performance-related anomalies that were fixed in the storage system such as: non-trivial implementation problems (e.g., limited randomness in the data placement decisions that created an artificial bottleneck, or unreasonable locking overheads at the manager). Similarly, the prediction mechanism helped us revisit assumptions about the middleware stack the storage system is implemented over (e.g., we have discovered the significant impact of the TCP connection initiation timeout of 3s in some scenarios); and highlighted shortcomings of the seeding process or incorrect assumptions about the deployment platform (e.g., we were ignoring platform heterogeneity). 

\vspace{+6pt}

\textbf{How to decide when to stop increasing the level of detail in the model and the complexity of system identification?}
We aim to model only the key interactions between system components. Modeling all system subcomponents and all their interactions in detail would be too complex. Such complexity could improve prediction accuracy, but would have significant drawbacks: significantly more complex model (as complex as the actual storage system and the underlying environment (e.g., network protocols, operating system buffers, scheduling), complex seeding process, lower scalability, and loss of the model generality. Further, the improvement in accuracy may not add much value (e.g., when the prediction mechanism is used to decide between system configurations). 
We followed a top-down approach: we started from a simple model and added more components or interactions' details until the accuracy of the all predictions was within 20\% of actual performance (and the median error was within 5\%) for the set of microbenchmarks we present in (\S\ref{subsec:microbenchmarks}). 

\vspace{+6pt}

\textbf {What is next?}
Our future work goes in several directions: (i) address the prediction shortcomings highlighted in \S\ref{sec:evaluation} -- besides the approaches already described, we also plan to enrich our system identification with data mining information \cite{iron-models}; (ii) fully automate the process to extract the behavior of the application based on log information and pass to the simulator (currently the logs already provides us the information needed, but the process is not fully automated), (iii) explore a richer space of configuration knobs; and (iv) evaluate the system using additional synthetic benchmarks and real applications.

\section*{Acknowledgements}

We would like to thank Hao Yang for his collaboration in MosaStore development; Elizeu Santos-Neto and Abdullah Gharaibeh for their collaboration in keeping the quality of the code via code reviews.

\bibliographystyle{splncs}
\bibliography{europar}

\end{document}